\documentclass[aps,prl,twocolumn, groupedaddress]{revtex4}
\usepackage[latin1]{inputenc}
\usepackage[english]{babel}
\usepackage{subfigure}
\usepackage{graphicx}
\usepackage{amssymb}
\usepackage{wasysym}
\usepackage{amscd}
\usepackage{color}
\def\gapx{\lower 2pt \hbox{$\buildrel>\over{\scriptstyle{\sim}}$\ }}
\def\lapx{\lower 2pt \hbox{$\buildrel<\over{\scriptstyle{\sim}}$\ }}
\begin{document}

\preprint{}

\title{Quantum spin liquid in a kagome lattice spin-1/2 XY model with four-site exchange}

\author{Long Dang, Stephen Inglis, and Roger G. Melko}
\affiliation{Department of Physics and Astronomy, University of Waterloo, 200 University Avenue, Ontario , Canada, N2L 3G1}

\date{\today}

\begin{abstract}
We study the ground state phase diagram of a two-dimensional kagome lattice spin-1/2 XY model ($J$) with a four-site ring exchange interaction ($K$) using quantum Monte Carlo simulations.
We find that the superfluid phase, existing in the regime of small ring exchange, undergoes a direct transition to a ${\mathbb{Z}}_2$ quantum spin liquid phase at $(K/J)_c \approx 22$, which is related to the phase proposed by Balents, Girvin and Fisher [{Phys. Rev. B,} {\bf 65}{ 224412 (2002)}].
The quantum phase transition between the superfluid and the spin liquid phase has exponents $z$ and $\nu$ falling in the 3D XY universality class, making it a candidate for an exotic XY* quantum critical point, mediated by the condensation of bosonic spinons.
\end{abstract}

\maketitle
The coveted quantum spin liquid state,
proposed more than three decades ago by Fazekas and Anderson \cite{Fazekas1974},
remains surprisingly elusive.  
Although recent experiments \cite{Robert2006,Shimizu2003,Coldea2001} provide tantalizing evidence of their existence,
the scarcity of spin liquid states in microscopic models belies their resistance to theoretical characterization.  Even
a basic unifying notion for which microscopic ingredients are required to promote spin liquid states in models is muddied.
Geometric frustration is the main suspect; recent Density-Matrix Renormalization Group measurements have championed the case for
a spin liquid as the long-debated groundstate of the kagome lattice Heisenberg antiferromagnet \cite{Steve}.  However, the apparent discovery of a
gapped spin liquid state in quantum Monte Carlo (QMC) simulations of the honeycomb lattice Hubbard model at half filling seems to contradict some long-held notions for where spin liquids might lie \cite{honeySL}.  The extensive theoretical framework \cite{Wen1991,Sachdev1992,Senthil2000,Motrunich2005}, developed over decades, continues to undergo refinement \cite{SB1,SB2,Cenke1} motivated by these and
other discoveries from large-scale computer simulations of a relatively small number of models.

Part of the difficulty in finding models that harbor spin liquids in two-dimensions (2D) and higher is the fact the important ingredient of frustration typically leads to the infamous sign problem and preclusion of the model from being studied by scalable QMC techniques.
A significant development in this front came in 2002, when Balents {\it et.~al.} \cite{Balents2002} proposed a sign-problem free Hamiltonian of spins on the kagome lattice which gives rise to a $\mathbb{Z}_2$ spin liquid phase.  Away from their exactly soluble point, the spin liquid appears to be robust, as
several models containing XY terms and constrained potentials have now shown evidence of this spin liquid ground state \cite{Sheng2005,Isakov2006a,Isakov2011}.
In this paper, we demonstrate that a $\mathbb{Z}_2$ spin liquid phase can also be stabilized in a model with competition between purely {\it kinetic} terms,
namely two-site XY and four-site ``ring'' exchange interactions.  This Hamiltonian is amenable to large-scale quantum Monte Carlo (QMC) studies, which allows us to characterize the spin liquid state -- and the quantum phase transition into this state -- in great detail.  We find that there 
exists a single direct quantum phase transition into the spin liquid from the superfluid phase, characterized by exponents $z$ and $\nu$ falling in the 3D XY universality class.  This makes the transition an 
 interesting candidate for an exotic XY* quantum critical point \cite{Chubukov1994a, Chubukov1994b,Isakov2005} caused by the condensation of bosonic spinons.

\begin{figure}
\includegraphics[scale=0.35,angle=0]{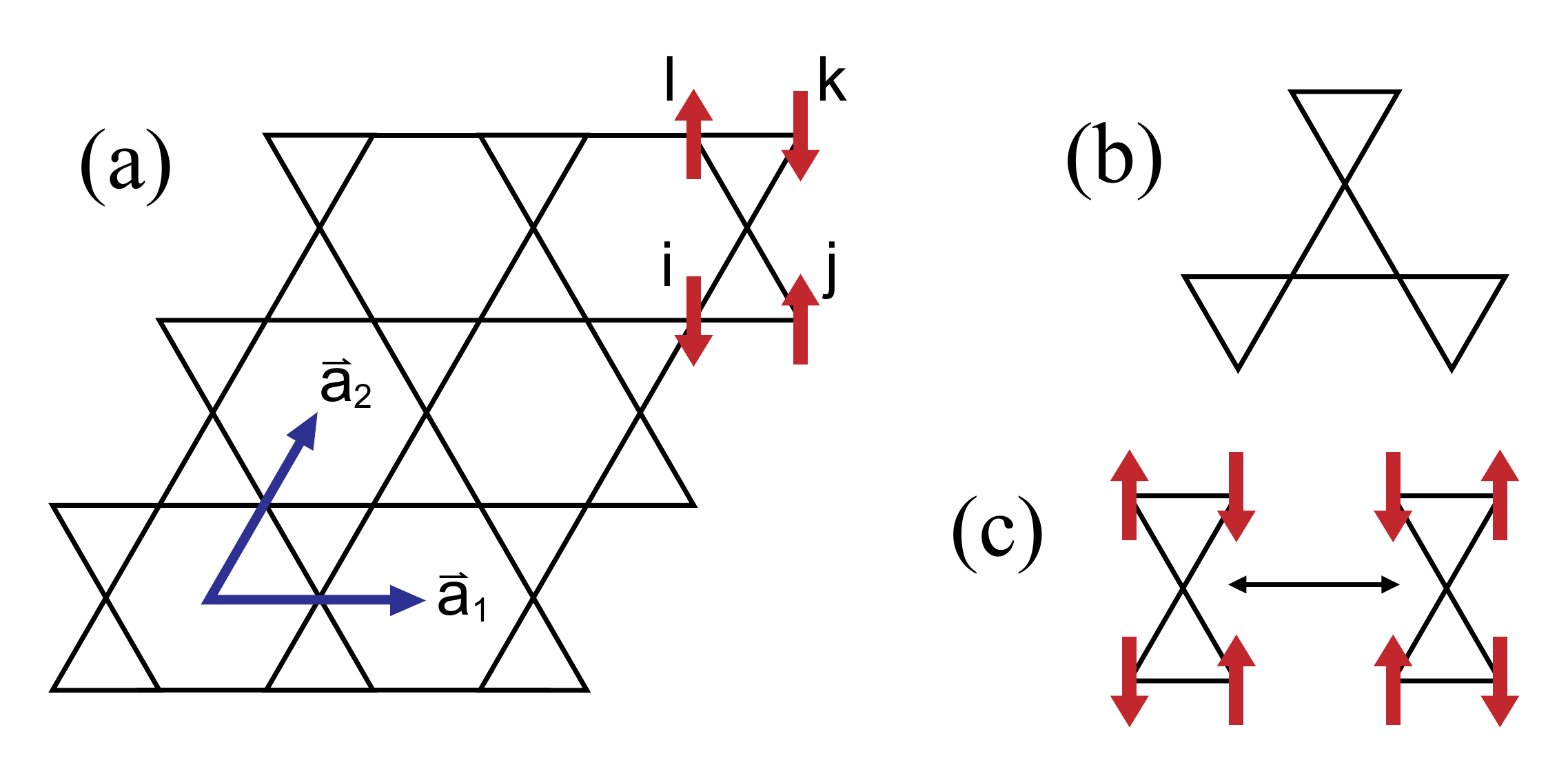}
\caption{(color online) (a) kagome lattice and a labeling convention for the indices of an operator $P_{ijkl}$.
Two primitive vectors $\vec{a_1}, \vec{a_2} $ are shown.(b) Three different orientations of the bow-tie plaquettes in the kagome lattice.
(c) The two spin plaquette configurations between which the $K$ term can act upon.
The up and down arrows correspond to up and down spins, respectively.} \label{f1}
\end{figure}

\begin{figure}
\includegraphics[scale=0.3,angle=0]{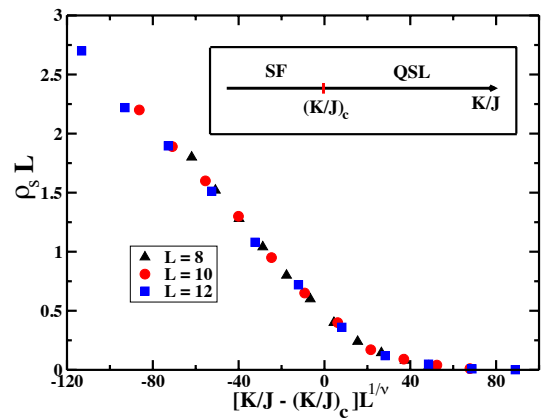}
\caption{(color online) Data collapse of the superfluid density $\rho_S$ for $\beta/L=2/J$, $(K/J)_c=21.8$, $z = 1$, and $\nu=0.6717$.
Inset: Schematic ground state phase diagram of the Hamiltonian (\ref{ham}) as a function of the 4-site ring exchange $K/J$.
The critical value of ring exchange $(K/J)_c=21.8$ separates the superfluid phase (SF) from the quantum spin liquid phase (QSL).  }\label{f3}
\end{figure}

We define
the standard J-K model by means of the U(1) symmetric spin-1/2 Hamiltonian:
\begin{equation}
H= -J\sum_{\langle ij\rangle} B_{ij} -  K\sum_{\langle ijkl\rangle} P_{ijkl}  \label{ham}
\end{equation}
where $J$ is the nearest neighbor coupling and $K$ is the four-site ring exchange.
The bond and plaquette operators are given by $B_{ij}= (S_i^+S_j^{-} + S_i^{-}S_j^+) = 2(S_i^{x}S_j^{x} + S_i^{y}S_j^{y})$ and  $P_{ijkl}= (S_i^+ S_j^{-}S_k^+ S_l^{-} + S_i^{-}S_j^+ S_k^{-}S_l^+)$.
The bond operator $B_{ij}$ represents the standard nearest-neighbor XY exchange interaction, where $\langle ij\rangle$ denotes a pair of nearest neighbor sites on a 2D kagome lattice.
The plaquette operators $P_{ijkl}$ (with labeling convention as in Fig.~\ref{f1}a) make up the purely XY part of the full four-site ring exchange interaction, where $\langle ijkl \rangle$ are sites on the corners of a bow-tie plaquette.
Similar J-K models have proven a fertile ground in the search for exotic phases and phase transitions in other lattices.
Recent proposals include Exciton Bose Liquids \cite{Arun1},
d-wave Bose liquids (DBL) \cite{DBL1}, and other exotic gapless Mott insulators \cite{GMI}.
The square-lattice XY ring-exchange model was the first example investigated as a candidate {\it deconfined} quantum critical point 
between a superfluid and Valence-Bond-Solid (VBS) phase, after QMC investigations pioneered by Sandvik {\it et. al.} \cite{Sandvik1999}
elucidated the ground-state phase diagram.
Interestingly, Buchler {\it et.~al.~}~\cite{Buchler2005} have recently presented the design of a ring-exchange interaction for cold bosonic atoms in 2D optical lattices, making the investigation of such models more than just a theoretical curiosity. 
Here, we examine the groundstate behavior of the kagome-lattice J-K model with QMC, restricting 
the sign of $K$ to be positive to avoid the sign problem.
In the limit of a vanishing ring exchange, e.g.  $K\rightarrow 0$, the J-K model reduces to the  spin {\it S} =1/2 XY model, 
where the hopping term $J$ drives the system into a  superfluid phase at low temperature.
The other limit, $K\rightarrow \infty$, leads to a pure ring-exchange model on the the kagome lattice, closely related to a model that has been explicitly shown to stabilize a spin liquid phase by Balents {\it et al.}~\cite{Balents2002}. In the presence of a fixed nearest neighbor hopping, one may expect  that the ring exchange term drives the system from the superfluid state at a small $K$ to a spin liquid state at an extremely large $K$.
Below, we show that the superfluid and spin liquid phases are separated by a single phase transition at $(K/J)_c=21.8$ (Fig.~\ref{f3}).

In order to study the groundstate phase diagram of the kagome J-K model using QMC, a variation of the stochastic series expansion (SSE) algorithm \cite{Melko2005} is used.  
We consider a finite size system mapped to a torus to have periodic boundaries in both directions, with size $\vec{L}_1 = n_1\vec{a}_1$ and $\vec{L}_2 = n_2\vec{a}_2$ (where $\vec{a}_1$ and $\vec{a}_2$ are primitive vectors and  $a_1=a_2=1$) as depicted in Fig.\ref{f1}a.
The total number of sites in the simulation cell is $N_s = n_1 \times n_2 \times 3$ sites. For simplicity, we use $n_1=n_2=L$ in our next discussion.
In order to perform the QMC, several essential modifications must be made to the SSE algorithm for the square-lattice J-K model using the conventional notations, i.e. J, C and K-vertices.  In the kagome case, the number of C and K vertices are the same in the SSE QMC scheme (see Ref.~\cite{Melko2005} Fig.~2). However, unlike the square lattice model, the J operators are only associated with the pair of two opposite spins highlighted with the arrows on the bow-ties plaquette as shown in Fig. \ref{f1}a.  In other words, the only valid bonds are ${ij}$ and ${kl}$ (whereas invalid bonds are $il$ and $jk$) for this particular bow-tie. As the results, the number of J vertices are reduced to 16, instead of 32 as in the case of square lattice. Technically, this leads to a unique and smaller set of directed loop equations, although the algorithm implementation in both cases are similar. Moreover, like the square-lattice model, simulation efficiency depends on the inclusion of special {\it multi-branch cluster} updates, which facilitate sampling of C $\rightarrow$ K \cite{Melko2005} vertices in the large-$K$ regime of the model. The square-lattice multi-branch cluster update must again be modified in a non-trivial way to accommodate the kagome lattice geometry, where the reduced set of unique J vertices reduces the number of possible vertex transformations.
Similar to the square lattice model, a single cluster can be constructed and flipped with probability $1$  (a Wolff update), or all clusters can be constructed and flipped independently with probability $1/2$ (a Swendsen-Wang update) \cite{Melko2005}. In the regime of large-$K$, we have observed that the Wolff-type update is more efficient to the Swendsen-Wang type.

In order to characterize the various phases in the model, we compute the spin stiffness, as well as the spin and plaquette structure factors.
The spin stiffness (superfluid density in the boson representation) is defined in terms of the energy response to a twist $\phi$ in the periodic boundary of the lattice:  $\rho_S = \frac{1}{N} \frac{\partial^2 E(\phi)}{\partial \phi^2} $.
Starting at small $K$ where the SSE algorithm is known to be ergodic,
the superfluid density $\rho_S$ decreases with increasing ring exchange, eventually vanishing at a critical value $(K/J)_c$.
At this point, the disappearance of superfluid density signals an insulating phase;
the smooth change of $\rho_S$ as a function of $K/J$ suggests a continuous quantum phase transition.
To examine the nature of this transition further, we apply scaling theory: $\rho_S = L^{-z}F_{\rho_S}(L^{1/\nu}(K_c-K)/J,\beta /L^z)$ \cite{rhoScale}, where $F_{\rho_S}$ is a universal scaling function, $L$ is the linear system size, $\beta$ is the inverse temperature or the imaginary time, $z$ is the dynamical critical exponent and $\nu$ is the correlation length exponent.
At a fixed ratio $\beta/L^z$ and sufficiently large inverse temperature, a plot of $\rho_SL^z$ as a function of $K/J$ should intersect at the critical point $(K/J)_c$.
Using this approach, we have identified the intersection at $(K/J)_c=21.8$ using $z=1$ and $\beta/L=2/J$ for different system sizes.
As illustrated in Fig. \ref{f3}, one may thereby determine the correlation length exponent $\nu$, by plotting $\rho_SL^z$ as a function of $[(K_c-K)/J]L^{1/\nu}$.
All the curves collapse into a single curve with $\nu=0.6717$ taken from a conventional 3D XY universality class \cite{Campostrini2006}. 
As discussed below, because of the condensation of bosonic spinons at the transition, it is predicted that this scaling behavior actually belongs to the different universality class called XY$^*$, in which an anomalous exponent $\eta$ is much larger than the one belonging to the 3D XY universality class, $\eta_{XY^*} \gg \eta_{XY}\approx 0.04$ \cite{Chubukov1994a,Chubukov1994b,Isakov2005,Grover2010}. 
This prediction could be addressed in future QMC studies.

\begin{figure}
\subfigure[]{
\includegraphics[height=0.38\columnwidth]{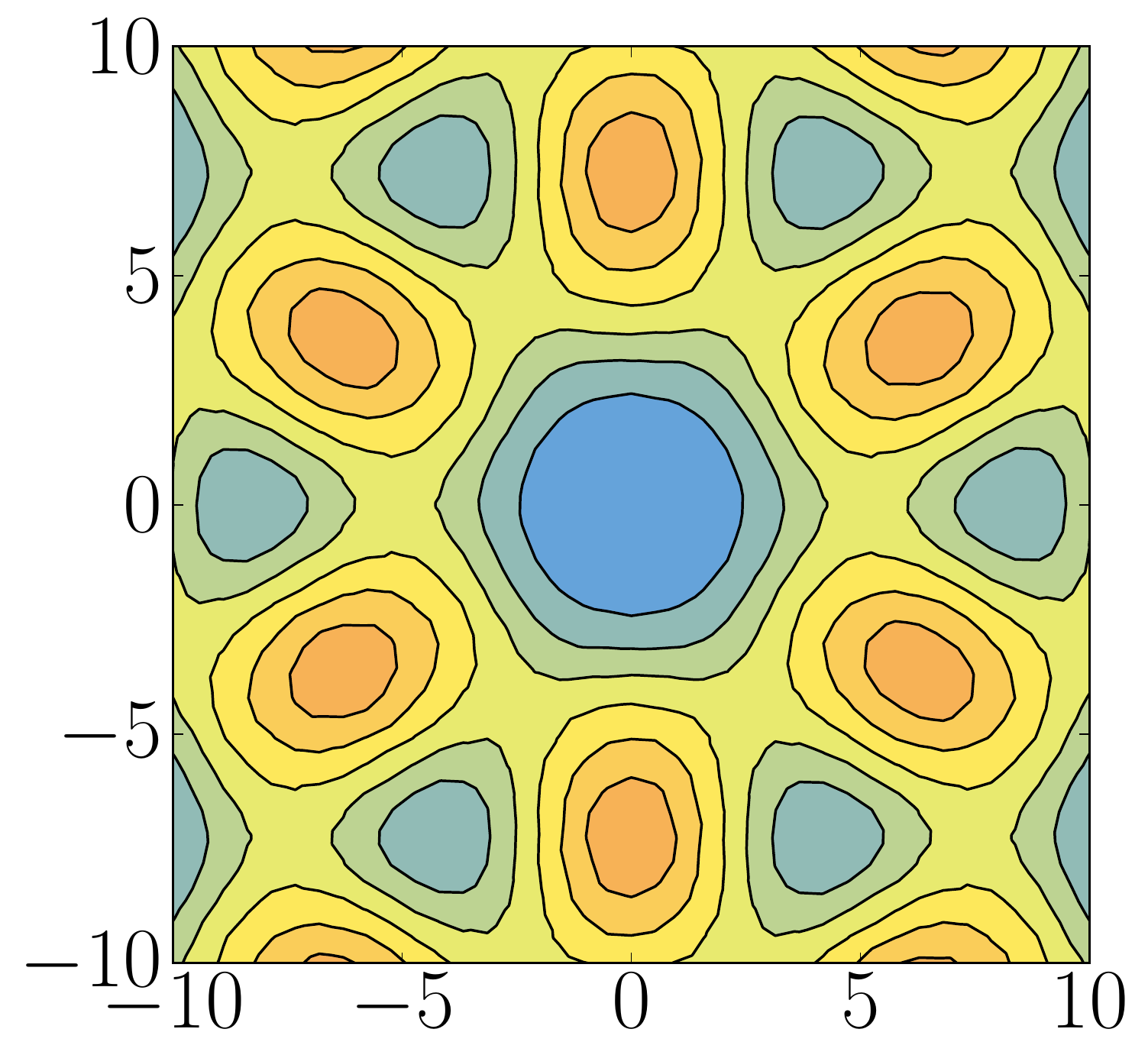}
}
\subfigure[]{
\includegraphics[height=0.38\columnwidth]{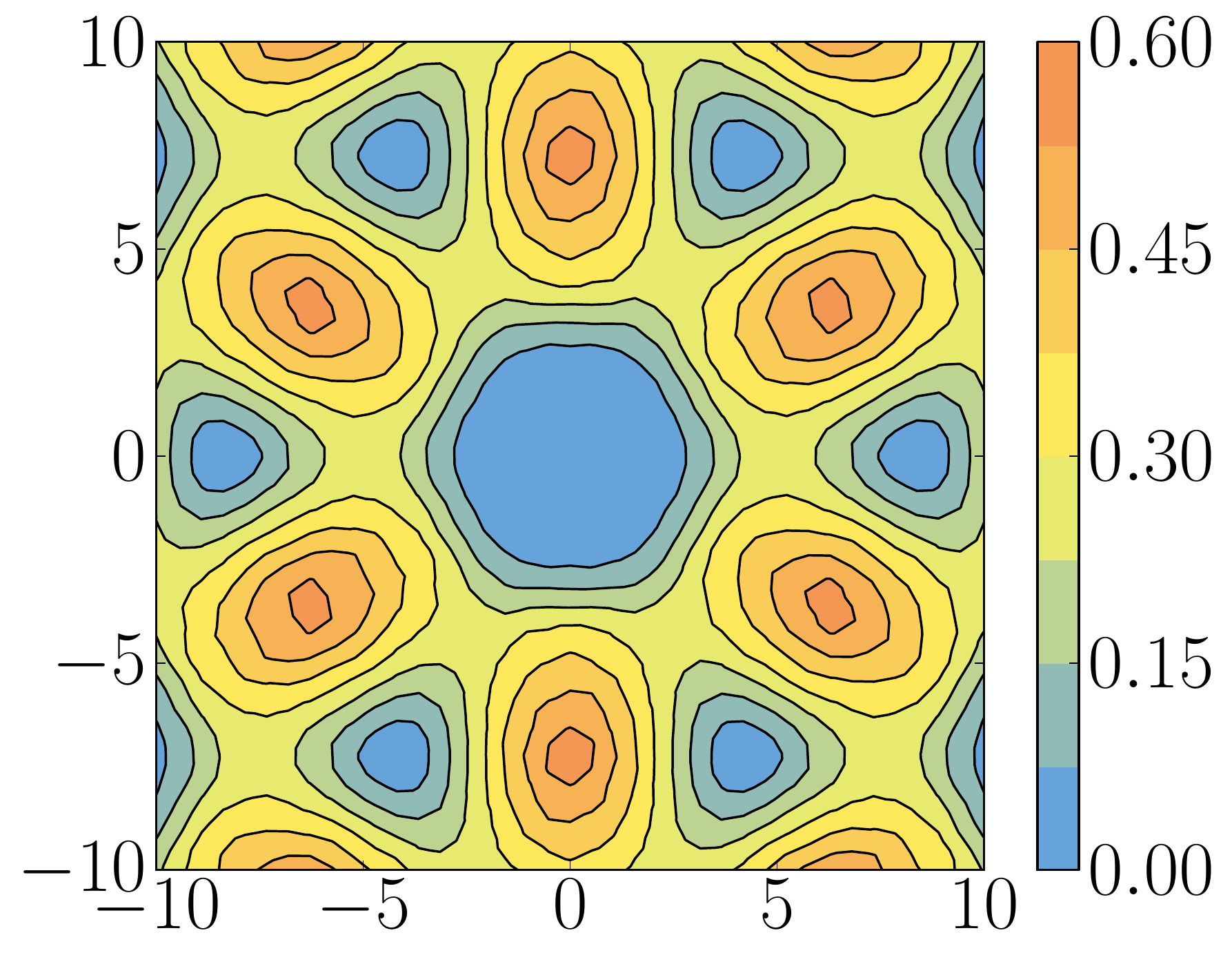}
}
\caption{
(color online) The spin-spin structure factor $S_s(\vec{\bf q})$ as a function of $(q_x,q_y)$ for the two states: (a) superfluid state with $K/J$ = 16 and (b) insulating state with $K/J$ = 26. ($L=12$) 
}\label{f4}
\end{figure}

\begin{figure}
\includegraphics[width=0.8\columnwidth]{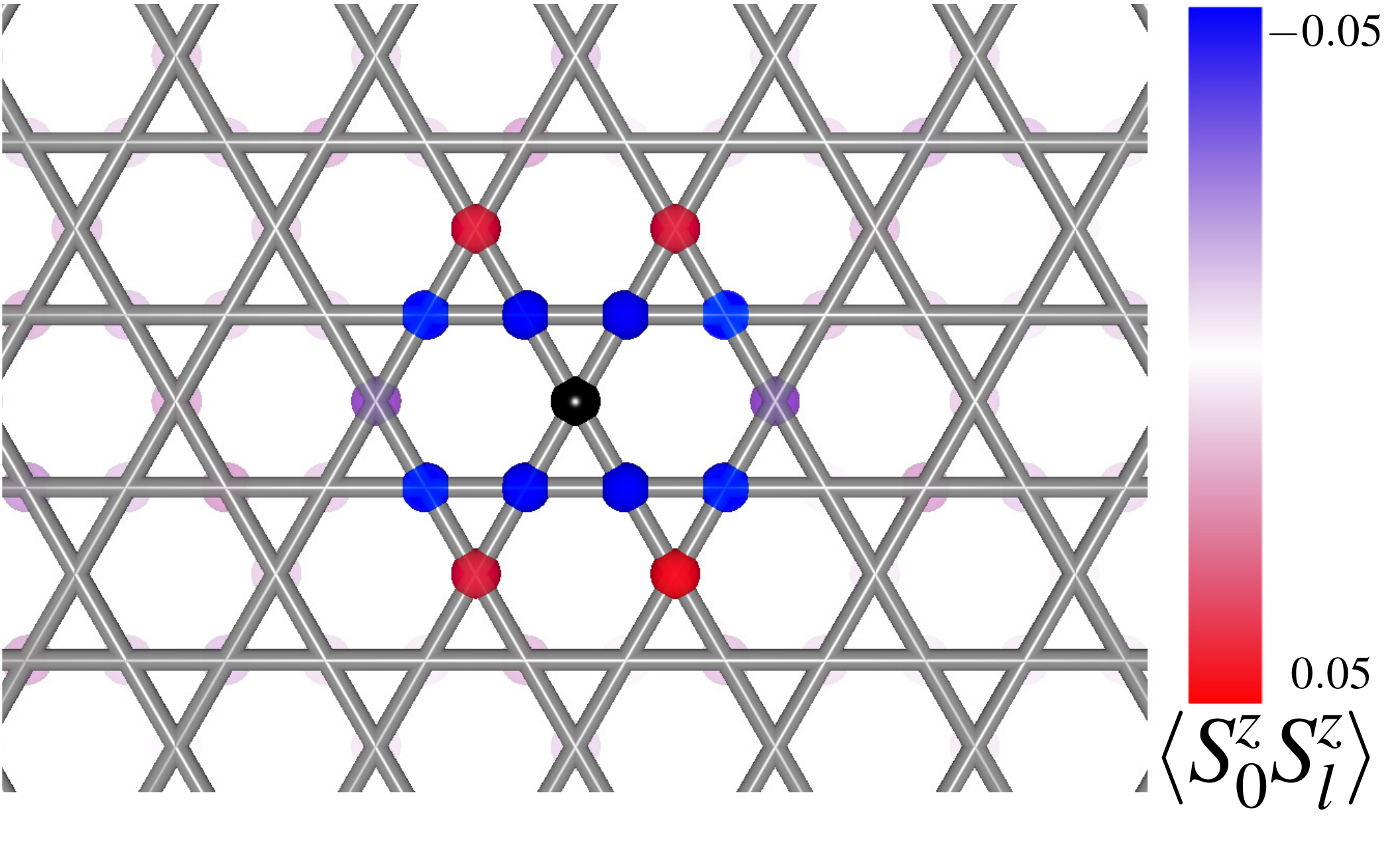}
\caption{(color online) The real-space image of the spin-spin correlation function, $\left< S^z_0 S^z_l \right>$. \label{f6}
The reference spin (black) at site $0$ is arbitrary.
The color scale is from red $\left< S^z_0 S^z_l \right> = 0.05$ to blue for $\left< S^z_0 S^z_l \right> = -0.05$ with increasing transparency as we approach  $\left< S^z_0 S^z_l \right> = 0$ (from either direction).
The picture shows most spins as transparent, or $\left< S^z_0 S^z_l \right> \approx 0$ for most $l$.
}
\end{figure}

\begin{figure}
\subfigure[]{
\includegraphics[height=0.38\columnwidth]{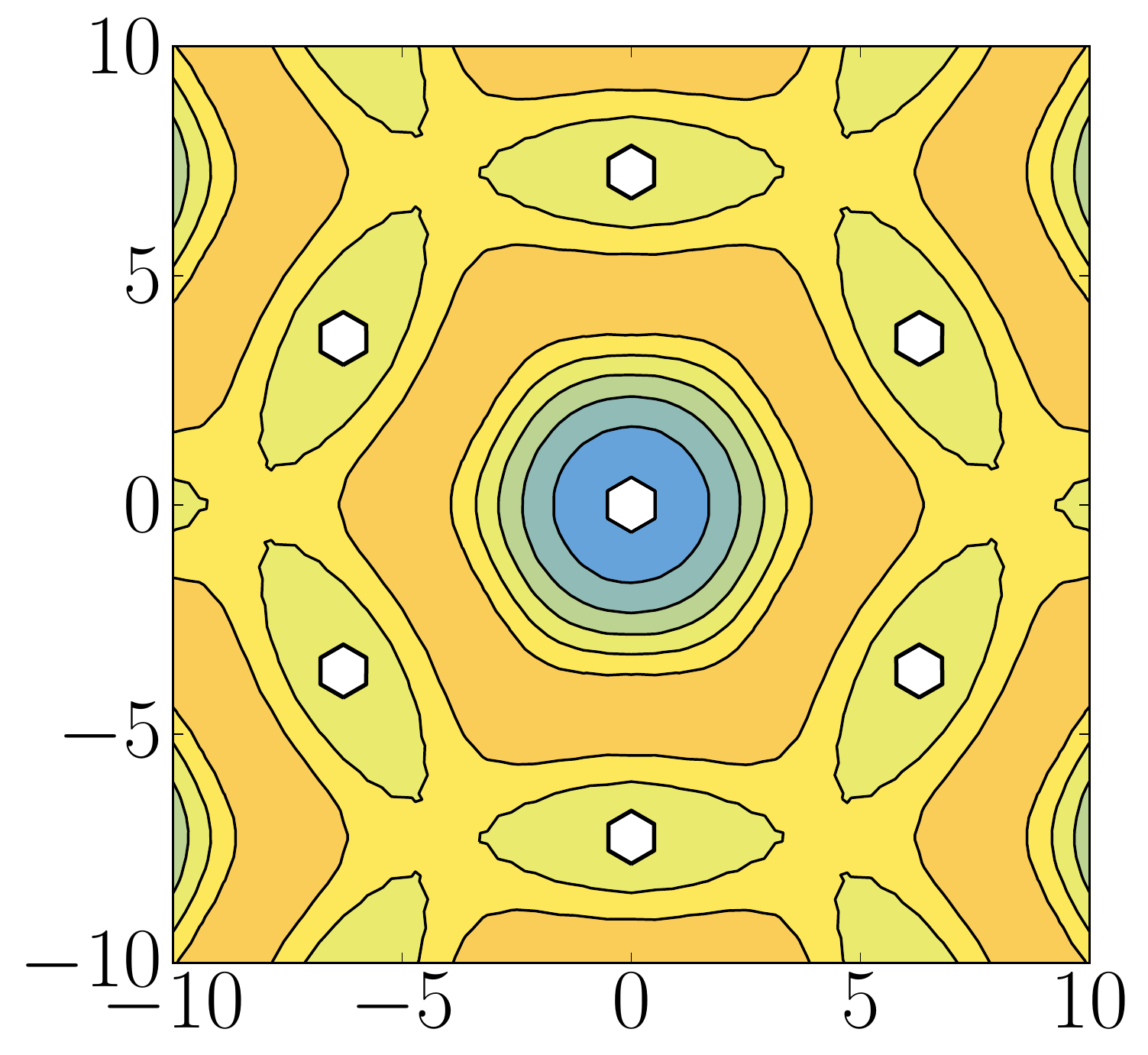}
}
\subfigure[]{
\includegraphics[height=0.38\columnwidth]{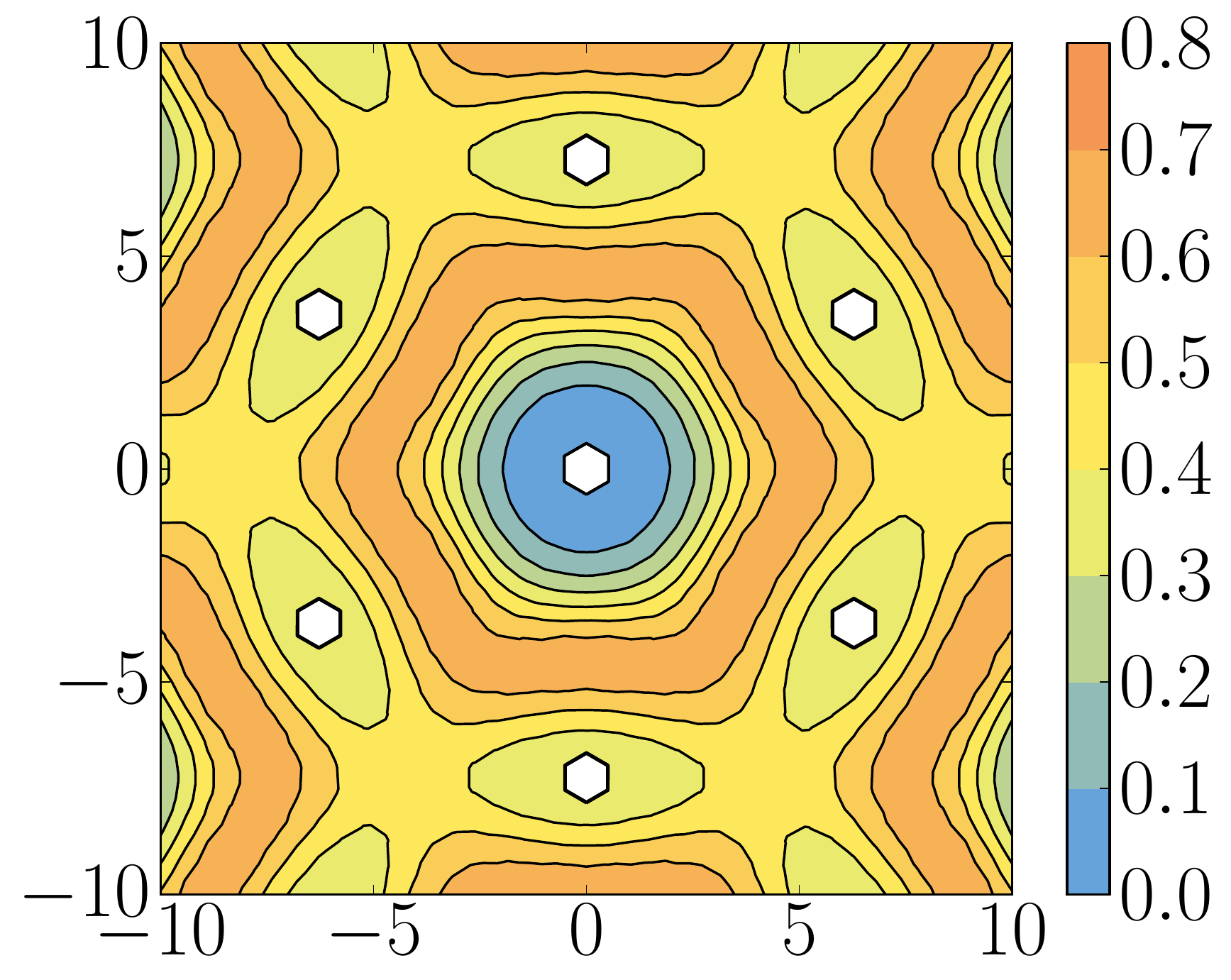}
}
\caption{
(color online) The plaquette-plaquette structure factor $S_p(\vec{\bf q})$ as a function of $(q_x,q_y)$ for the two states: (a) superfluid state with $K/J$ = 16 and (b) insulating state with $K/J$ = 26. ($L=12$) 
}\label{f5}
\end{figure}

The adherence of the superfluid-insulating transition to this universality class lends credence to the hypothesis that the insulating
phase is in fact a $\mathbb{Z}_2$ spin liquid, as seen in previous studies \cite{Isakov2006a,Isakov2011}.
Further evidence can be obtained by examining the z-component of the spin-spin correlation functions, which we average over the propagated states, i.e. $\langle S_k^z S_l^z \rangle = \frac{1}{4} \left \langle \frac{1}{n}\sum_{p=1}^{n-1} \sigma_k^{z}(p) \sigma_l^{z}(p)   \right \rangle \label{corr}$, 
where {\it n} is the number of non-identity operators in the SSE operator list, the z-component $S_{k,l}^z=(1/2)\sigma_{k,l}^z$ (with $\sigma_{k,l}^z=\pm1$) \cite{Melko2005}. 
These are used to construct the spin structure factor $S_s(q_x,q_y) = \frac{1}{N}\sum_{k,l}e^{i({\bf r}_k-{\bf r}_l).{\bf q}}  \langle S_k^z S_l^z \rangle$, where {\it k} and {\it l} are lattice sites; ${\bf r}_i=(x_i,y_i)$ is the lattice coordinate (with lattice spacing chosen as unity) and ${\bf q}=(q_x,q_y)$ is the wave vector.
In order to examine the characteristics of the insulating phase, we investigate the spin structure factor $S_s({\bf q})$ as illustrated in Fig.~\ref{f4}.
The spin structure factor looks similar in both the superfluid state and the insulating phase corresponding to $K/J = 16$ and $K/J = 26$, respectively.
At any ${\bf q}$ vector, there are no extensively-scaling Bragg peaks (required for long-range order) in the insulating state.
The non-extensive features persisting in large system sizes, e.g.~$L = 12$, reflect the existence of short-range correlations.
This evidence rules out a possible charge density wave with a broken translation symmetry.
It is also useful to visualize the real space structure of the spin-spin correlation function in the insulating state: $K/J = 26$ is shown in Fig.~\ref{f6}.
The real-space picture shows a local configuration that reduces nearest neighbor hopping energy ($J$) while the reference spin (black) is also able to participate in exchange with all four plaquettes it is a part of.
This configuration is in line with what one would expect to find through performing local perturbation theory.

 In Fig.~\ref{f5}, we present the plaquette structure factor $S_p(q_x,q_y) = \frac{1}{N} \sum_{a,b} e^{i({\bf r}_a-{\bf r}_b).{\bf q}} \langle P_a P_b \rangle$, where $P_a$ is the plaquette operator with the plaquette subscript $\it a$ defined in Fig.~\ref{f1}c. This characterizes the modulations of the plaquette operator expectation value $\langle P_{ijkl} \rangle$.
Similar to the spin structure factor, peaks in the plaquette structure factor indicate a VBS phase only if they survive in the thermodynamic limit.
Clearly, the plaquette structure factor $S_p({\bf q})$ shows no evidence of any extensively scaling peaks~\footnote{The $\bf{q}=(0,0)$ peaks and their reflections only represent a finite density of plaquettes, not ordering} at any ${\bf q}$ vector, ruling out the possibility of having a VBS for the insulating phase. 

In summary, we have studied the ground state phase diagram of the kagome lattice spin 1/2 XY model with four-site ring exchange model using SSE-QMC calculations.
We find a continuous quantum phase transition at $(K/J)_c \approx 22$ from the superfluid phase to a quantum spin liquid.
Finite-size scaling studies confirm to high accuracy that the quantum critical point has a dynamical exponent $z = 1$ and a correlation length exponent $\nu = 0.6717$ falling into the 3D XY universality class.
Above this transition, we observe a large region of a gapped spin liquid phase, which dominates the phase diagram.  Spin and plaquette correlation functions are observed to have no long-range order over the entire Brillouin zone.

This spin liquid phase should be adiabatically connected to the $\mathbb{Z}_2$ spin liquid phase proposed by Balents {\it et. al.}, which is demonstrated to
be stable in other related models \cite{Sheng2005,Isakov2006a,Isakov2011}.  
This is supported by the similarity in the {\bf q}-space spin structure factor between our ring-exchange system and these models.
In addition, the fact that the superfluid-insulator transition is observed to be in 
the 3D XY universality class also provides strong evidence that the featureless insulator phase is in fact a spin liquid.  
In future studies, it would be interesting to investigate the finite temperature phase diagram accessible with SSE-QMC.  In particular,
smoking-gun evidence of the emergent ${\mathbb {Z}}_2$ topological order should be possible using Renyi entropy 
measurements \cite{XXZ}, 
however due to the significantly stronger critical coupling in the J-K model, these studies may require more computational resources than the model in Ref.~\cite{Isakov2011}.

This model provides a novel route to studying exotic criticality, since
 the scaling behavior of the quantum phase transition may actually belong to an exotic XY$^*$ universality rising from the 
 condensation of bosonic spinons.
Measurements of the anomalous dimension exponent would be sufficient to confirm this novel behavior, especially since the numerical
value expected to be much larger (of order $\eta = 1.37$ \cite{Isakov2005}) than $\eta$ at a conventional transition.
Just as interestingly, extensions of this model with symmetry breaking interactions are expected to give rise to a measurable vison-confinement, which should also be manifest as an anomalously large (but unique) value of $\eta$.  Other interesting phases, such as 
exotic supersolid phases, may also be present in straightforward extensions of the model, such as with the application of a uniform external field.

Our work has identified a broad new class of XY ring-exchange model which supports an extended region of quantum spin liquid, amenable
to large-scale QMC simulations without the sign problem.  The rise of the spin liquid due to the competition of purely kinetic interactions,
which do not contain conventional geometric frustration,
marks significant progress in the understanding of what ingredients are necessary to promote spin liquids in realistic models \cite{Buchler2005}.  The abundance of possible extensions to this model which are likely to show exotic phases or phase transitions suggests that large-scale simulation work on 
kagome-lattice J-K models has only yet begun.

We are grateful to S. Isakov,  T. Senthil and C. Xu for fruitful discussions.
R.G.M. would like to acknowledge the support and hospitality of Microsoft Station Q.
This work was made possible by the computing facilities of SHARCNET.
Support was provided by NSERC of Canada. 
\bibliography{kagome}{}

\end{document}